\documentclass{llncs}
\usepackage{listings}
\usepackage{enumitem}
\usepackage{booktabs}
\usepackage{multirow}
\usepackage{makecell}
\usepackage{siunitx}
\usepackage{graphicx}
\usepackage{epsfig,epstopdf}

\usepackage{amsmath}
\newcommand{\isep}{\mathrel{{.}\,{.}}\nobreak}

\begin{document}
\title{Neural Feature Embedding for User Response Prediction in Real-Time Bidding (RTB)}

\author{Enno Shioji\inst{1}\inst{2} \and Masayuki Arai\inst{2}}

\institute{
$^1$Adform A/S, Wildersgade 10B, DK-1408 Copenhagen, Denmark\\
$^2$Graduate School of Science and Engineering, Teikyo University, Japan\\\email{$^1$Enno.Shioji@adform.com, $^2$arai@ics.teikyo-u.ac.jp}}
\maketitle

\begin{abstract}
In the area of ad-targeting, predicting user responses is essential for many applications such as Real-Time Bidding (RTB). Many of the features available in this domain are sparse categorical features. This presents a challenge especially when the user responses to be predicted are rare, because each feature will only have very few positive examples. Recently, neural embedding techniques such as word2vec which learn distributed representations of words using occurrence statistics in the corpus have been shown to be effective in many Natural Language Processing tasks. In this paper, we use real-world data set to show that a similar technique can be used to learn distributed representations of features from users’ web history, and that such representations can be used to improve the accuracy of commonly used models for predicting rare user responses.
\end{abstract}

\section{Introduction}\label{sec:intro}
Predicting the probability of user response such as click, conversion etc. given an ad-impression is crucial for many advertisement applications, such as Real-Time Bidding (RTB). Because of its efficiency, linear models such as logistic regression are the most widely used for this purpose\cite{rtbreview}. The models are commonly trained on sparse categorical features such as user agent, IDs of visited websites etc., which are encoded as sparse binary features via one-hot encoding\cite{rtbreview}. One of the prominent problem with these models is the sparsity of data. Especially when feature interaction is used, the feature representation becomes extremely sparse, making it difficult to exploit the features efficiently. Moreover, traditionally the industry has focused on predicting clicks, but recently the focus has shifted to optimizing for other, much rarer user responses like conversions, which exacerbates this problem\cite{dalessandro:proxy}. We refer to this problem as feature sparsity problem. \par

A similar issue has been recognized in Natural Language Processing (NLP)\cite{neuralproblan}. Many mainstream models rely on bag-of-words representation, which suffers from the same issue outlined above. Recently, neural embedding techniques known as word2vec, paragraph2vec etc. that map words and documents into low-dimensional vector space has been shown to yield state-of-the-art results in various NLP tasks\cite{mikolov:wr2,paravec}. In this approach, occurrence statistics in the corpus is used learn distributed word representations that are much more amenable to generalization. \par

In this paper, we use real-world data set to show that a similar technique can be applied to user response prediction in RTB. Similar to the situation in Natural Language Processing, a large amount of user web history can be used to learn high quality feature representations, which can then be used to predict (rare) user responses. The technique was shown to improve the accuracy of commonly used models, especially when labeled data was scarce.

\section{Related Works}\label{sec:related}
Various methods have been employed to address the feature sparsity problem. For example, higher order category information derived from human annotation, or from the data via unsupervised methods such as topic modelling, clustering etc.\cite{zhangdeeplearning,ipinyou} has been used to improve generalization. Other techniques such as counting features can also help by allowing rare features to contribute jointly\cite{fbhe}. \par

Another category of solutions involve embedding sparse categorical features into low-dimensional vector space. Various feature transformation methods that yield dense features has been investigated in conjunction with deep neural networks resulting in improvement over major state-of-the-art models\cite{zhangdeeplearning}. Zhang et al. also investigated a framework they refer to as implicit look-alike modelling, in which entities like users, web-pages, ads etc. are mapped into a latent vector space using both general web browsing behavior and ad response behaviour data\cite{implicitlal}. \par

In this paper, we report initial results of applying a feature transformation technique similar to neural word embedding to user response prediction in RTB. The technique has been successfully applied to other domains, such as product recommendation \cite{content2vec,item2vec}. The technique shares the benefits of its counterpart in NLP, such as the ability to encode feature sequences, the ability to incrementally update the embeddings with new data, and the availability of numerous improvements and extensions that have been developed since its advent. The result opens up exciting opportunities to apply techniques that have been successfully used with neural word embeddings, such as deep neural networks.

\section{Neural Feature Embedding for User Response Prediction}\label{sec:neurem}
We first provide a brief overview of the neural word embedding technique developed by Mikolov et al\cite{mikolov:wr}. We consider one of its simplest form, the Continuous Bag-of-Word Model (CBOW) with a single context window. Given a word $t$ in the corpus and the previous word $c$, we parametrise $\theta$ such that the conditional probabilities $p(t|c;\theta)$ is maximized for the corpus. $p(t|c;\theta)$ can be modelled using soft-max as follows:

\begin{equation}
p(t|c;\theta) = \frac{e^{v_c \cdot v_t}}{\sum_{t'\in V}e^{v_c \cdot v_{t'}}}
\end{equation}

\noindent where $v_t$ and $v_c \in R^n$ are vector representations for $t$ and $c$, and $V$ is the set of all vocabulary. $n$ is a hyper-parameter that determines the size of the embedding, and is chosen empirically. Note that we use a distinct representation for target and context, following the literature. This objective is straightforward but expensive to calculate. To alleviate this problem, a technique called negative sampling\cite{mikolov:wr} is used, wherein random pairs of $(t, c)$ is sampled from the corpus, assuming they are wrong. This yields the following objective: 

\begin{equation}
\arg\max_\theta \sum_{(t,c)\in D} \log \frac{1}{1+e^{- v_c \cdot v_t}} +
\sum_{(t,c) \in D'} \log (\frac{1}{1+e^{v_c \cdot v_t}})
\end{equation}

\noindent where $D$ is the set of all target-context pair in the corpus and $D'$ are randomly generated $(t,c)$ pairs. The objective is now cheap to calculate. \par

In this paper, we consider a dataset consisting of ad impressions. When an ad is shown to a user, some of the browsing history of that user is available as sequence of content IDs. It is thus relatively straightforward to apply techniques such as CBOW\cite{mikolov:wr}, skip-gram\cite{mikolov:wr} etc. to this data. For this experiment we chose to discard the sequence of the content IDs and only use the co-occurrence information. More specifically, we generated our positive $(t,c)$ pairs by randomly sampling content IDs from the set of content IDs the user had consumed at the time of the impression, and our negative pairs randomly from the corpus. It is known that the probability distribution of such sampling influences the quality of the embeddings\cite{mikolov:wr2}, but we used a uniform distribution for this initial experiment. We then used the resulting content embeddings as features in our user response model, for which we use logistic regression.\par

\section{Experiment and Discussion}\label{sec:Experiment}
\subsection{Dataset}
We used a real-world RTB dataset provided by Adform. Each record in the data corresponds to an ad-impression, and is ordered chronologically. The record consists of a binary label that indicates whether the user subsequently clicked the ad (\verb|click|), and a set of content IDs (\verb|content_ids|) the user had consumed in the past 30 days, up to the time of the impression. The data was taken from Adform's impression logs of July 2016. Records for which no \verb|content_ids| were available were filtered out. Further, negative examples were down-sampled at a rate of 0.01 as the data is extremely imbalanced. After the down-sampling, there were 5.0M examples in total, with 1.1M positive examples. There were 891K distinct content IDs. A newer, larger version of the dataset with additional fields has been published \cite{adformdata}. The \verb|content_ids| correspond to feature \verb|c9| in this dataset.

\subsection{Experiment Protocol}
The experiment consisted of an unsupervised stage and a supervised stage.  \par
\noindent \textbf{Unsupervised stage.} Content embeddings were learned from \verb|content_ids| as described above. I.e. the \verb|click| field was discarded and not used for this stage. Out of the 5.0M data instances, the oldest 4.0M were used for this stage. We trained the embeddings with varying embedding sizes $n$ ($2^{k\in[1\isep 7]}$). Tensorflow\cite{tensorflow} was used to implement this stage.

\noindent \textbf{Supervised stage.} In the supervised stage, binary classifiers that predicts \verb|click| were trained, using different features (see below). For all experiments, Logistic Regression with L2 normalization was used. Out of the remaining 1.0M data instances, the newest 30\% (300K) were held-out as validation dataset. The training was done with varying amounts of data (0.3K, 1K, 10K, 100K) that were randomly sampled from the remaining data (700K). To evaluate the performance of the models, area under the ROC curve (AUC) was used, which is a commonly used metric for evaluating user response prediction models in RTB\cite{rtbreview}. Grid search was performed with varying regularization strength ($10^{k\in[-2\isep 1]}$) and embedding size, and the best result was used as measurement. scikit-learn\cite{scikit-learn} was used for the implementation. Below is the list of features we compared:\\
\begin{itemize}[leftmargin=15.5mm]
  \item [SB:] Sparse Binary. \verb|content_ids| were encoded as Sparse Binary features via one-hot encoding. This is our baseline.
  \item [DR:] Distributed Representation. Each dimension of the resulting embeddings were scaled by its maximum absolute value. For each \verb|content_id| in \verb|content_ids|, the corresponding embedding was looked up and the mean of the embeddings were used as the feature vector. The resulting feature vector had thus the same length $n$ as the embeddings.
  \item [SB+DR:] Sparse Binary and Distributed Representation. The feature vector of SB and DR were concatenated.
\end{itemize}

\subsection{Performance Comparison and Discussion}
Table \ref{t:overall} shows the best results obtained for each condition using the aforementioned grid-search. The results of SB+DR and DR is compared against SB (our baseline). DR outperforms SB when training data is scarce. SB+DR outperforms SB in all conditions, especially stronger when training data is scarcer. This is likely because when training data is scarce, the sparsity issue is more acute and thus the ability to generalize across features has a larger effect. However, when large amount of data is available, the lower-dimensional feature representation of DR likely limits the degree of differentiation between individual content IDs. When SB and DR is concatenated, both advantages can be preserved. 

\begin{table}
\caption{AUC performance of click prediction} \label{t:overall}
\begin{center}
\setlength\tabcolsep{10pt}
  \begin{tabular}{l|cccc|S}
    \toprule
    \multirow{2}{*}{Training Data} &
      \multicolumn{2}{c}{SB+DR} &
      \multicolumn{2}{c}{DR} &
      \multicolumn{1}{c}{SB} \\
      & {AUC} & {AUC Lift} & {AUC} & {AUC Lift} & {AUC}  \\
      \midrule
      0.3K   &     60.93\% &   4.33\%  & \bfseries 61.53\% &   4.90\%  &   56.60\% \\
        1K   &     \bfseries 65.15\% &   3.39\%  & 64.20\% &   2.44\%  &   61.76\% \\
        10K  &     \bfseries 67.42\% &   2.87\%  & 65.47\% &   0.92\%  &   64.55\% \\
        100K &     \bfseries 69.65\% &   1.34\%  & 65.94\% &   -2.37\% &   68.31\% \\
    \bottomrule
  \end{tabular}
\end{center}
\end{table}

\begin{figure}[h]
\centering
\includegraphics[scale=0.28]{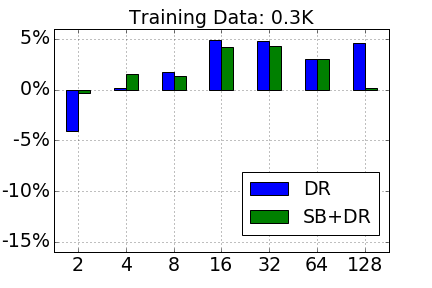}
\includegraphics[scale=0.28]{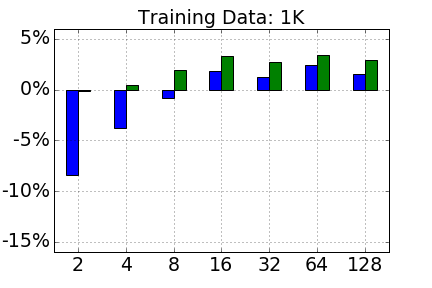}
\includegraphics[scale=0.28]{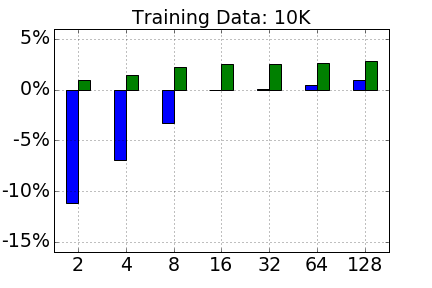}
\includegraphics[scale=0.28]{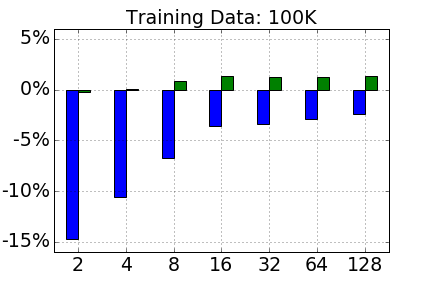}
\caption{AUC Performance difference from baseline against embedding size $n$}\label{fig:n_to_auc_lift}
\end{figure}

Figure \ref{fig:n_to_auc_lift} shows the difference in AUC from the SB baseline for DR and SB+DR, for varying embedding sizes ($n$). Increasing $n$ improves AUC, but the return diminishes after about 16 dimensions. 

\section{Conclusion}\label{sec:Conclusion}

In this paper, we reported initial results of applying neural feature embedding technique to user response prediction in RTB, using a real-world dataset. To our best knowledge, this is the first time this technique was applied to this problem. We have demonstrated that the technique can improve performance of commonly used model in the industry, especially when labeled data is scarce and when thus the feature sparsity problem is most acute. The fact that large amount of data can readily be used for training of the feature embeddings, and that the commonly used logistic regression can be used at prediction time make the result ideal for industrial implementation. \par
The result also opens up exciting opportunities to apply improvements and techniques that have been developed around neural word embeddings, such as incorporating global context, using multiple representations per word\cite{huang:improvewv}, optimizing the embeddings for a specific supervised task using target labels\cite{reembed}, using a global log-bilinear regression instead of the earlier local context window methods\cite{glove}, applying deep neural networks on the embeddings etc.
\bibliographystyle{splncs03}
\bibliography{biblio}
\end{document}